\documentstyle[aps,floats,epsf]{revtex}  

\draft
\begin{document}

\twocolumn[\hsize\textwidth\columnwidth\hsize\csname
@twocolumnfalse\endcsname

\title{Finite-temperature properties
of Pb(Zr$_{1-x}$Ti$_{x}$)O$_{3}$ alloys from first principles}
 
\author{L. Bellaiche$^{1}$, Alberto Garcia$^{2}$ and David Vanderbilt$^{3}$}

\address{$^{1}$ Physics Department,\\
                University of Arkansas, Fayetteville, Arkansas 72701, USA\\
         $^{2}$Departamento de Fisica Aplicada II, \\ Universidad 
del Pais Vasco, Apartado 644,
             48080 Bilbao, Spain \\
        $^{3}$  Center for Materials Theory,
                   Department of Physics and Astronomy,\\
         Rutgers University, Piscataway, New Jersey 08855-0849, USA}

\date{April 21, 2000}

\maketitle

\begin{abstract}
A first-principles-derived approach is developed to study
finite-temperature properties of Pb(Zr$_{1-x}$Ti$_{x}$)O$_{3}$
(PZT) solid solutions near the morphotropic phase boundary (MPB).
Structural and piezoelectric predictions are in excellent agreement
with experimental data and direct first-principles results.  A
low-temperature monoclinic phase is confirmed to exist, and is
demonstrated to act as a bridge between the well-known tetragonal
and rhombohedral phases delimiting the MPB.  A successful
explanation for the large piezoelectricity found in PZT ceramics
is also provided.
\end{abstract}


\vskip2pc]

\narrowtext

Ferroelectric perovskite A(B$'$B$''$)O$_{3}$ alloys are of growing
importance for a variety of device applications
\cite{Uchino,Akbas2}, and are also of great current fundamental
interest since little is known about the effects responsible for their
anomalous properties.  A good example of an A(B$'$,B$''$)O$_{3}$
solid solution that is of both fundamental and technological importance
is the Pb(Zr$_{1-x}$Ti$_{x}$)O$_{3}$ system.  Usually denoted as PZT,
this mixed-cation alloy is currently in widespread use in piezoelectric
transducers and actuators \cite{Uchino}.  Its phase diagram exhibits a
morphotropic phase boundary (MPB) separating a region with a
tetragonal ground state ($x$ $ >$ 0.52) from a region with
rhombohedral symmetry ($x$ $<$ 0.45) \cite{Lines}.

High piezoelectric response is experimentally found in {\it
ceramics} of PZT around the MPB.  The origins of this large
piezoelectric response are unclear.  On the one hand,
semi-empirical simulations predict that the large experimental
value of the $d_{33}$ piezoelectric coefficient results mainly from
the large value of $d_{33}$ that a {\it single-crystal} PZT would
exhibit \cite{Du1}. On the other hand, recent first-principles
calculations \cite{Szabo1,LaurentDavid1} have found that the
$d_{33}$ coefficient of a tetragonal single crystal of
Pb(Zr$_{0.5}$Ti$_{0.5}$)O$_{3}$ are estimated to be {\it three times
smaller} than the experimental value obtained for ceramics at low
temperature.

Furthermore, recent synchrotron x-ray powder diffraction studies have
revealed the existence of an unexpected low-temperature monoclinic
phase of PZT at $x$=0.48 \cite{Noheda1}, which
implies that the phase diagram of PZT is more complex than
previously thought.  This monoclinic phase may act as a
second-order transitional bridge between the tetragonal phase,
for which the electrical polarization $\bf{P}$ lies along the
pseudo-cubic [001] direction, and the rhombohedral phase, for
which $\bf{P}$  is along the pseudo-cubic [111] direction.  If this
is indeed the case, the polarization of the monoclinic phase
continuously rotates as the composition $x$ decreases in the MPB
region \cite{Noheda1}.  Such a continuous rotation has yet to be observed.

Obviously, accurate simulations are needed to understand the
properties of perovskite alloys in general, and of PZT in
particular.  Since the beginning of the present decade,
first-principles methods have emerged as a powerful tool for
investigating properties of ferroelectric systems theoretically
(see \cite{Szabo1,LaurentDavid1,Davidreview,LaurentJorgeDavidPRB}
and references therein).  However, these methods are essentially
restricted to the study of the zero-temperature properties of small
cells, while accurate and interesting predictions of alloy
properties would require calculations on much larger cells at
finite temperature.  Ideally one desires a
computational scheme with the capability of predicting the properties
of ``real'' perovskite alloy systems at finite temperature, with
the accuracy of the first-principles methods.

The purpose of this letter is to demonstrate that it is possible to
develop such a scheme, and to apply it to study the
finite-temperature behavior of PZT in the vicinity of the MPB.
Remarkably, we find that the existence of an intermediate
monoclinic phase emerges naturally from this approach.
 Moreover, the theory provides a novel and successful
explanation for the large piezoelectric response of PZT near the
MPB, thereby explaining and resolving the previous theoretical
difficulties in obtaining agreement with the known experimental
values of the piezoelectric coefficients.

Our scheme is based on the construction of an effective Hamiltonian
from first-principles calculations.  A {\it ferroelectric}
effective Hamiltonian \cite{ZhongDavid} must include the
ferroelectric local soft mode and the strain variables, since
ferroelectric transitions are accompanied by a softening of a polar
phonon mode and by the appearance of a strain.  An {\it alloy}
effective
Hamiltonian must also include the compositional degrees of
freedom.  We propose to incorporate all such degrees of
freedom by writing the total energy $E$ as a sum of two energies,
\begin{eqnarray}
   E (\{ { \bf u_{\it i}} \}, \{ { \bf v_{\it i}} \}, \eta_{\it H} &,&
      \{ \sigma_{\it j} \}) =
   E_{\rm ave} (\{ { \bf u_{\it i}} \}, \{ { \bf v_{\it i}} \}, \eta_{\it H})
   \nonumber \\
   &&\quad+~E_{\rm loc} (\{ { \bf u_{\it i}} \},\{ { \bf v_{\it i}} \},
   \{ \sigma_{\it j} \}) \;\;,
\end{eqnarray}
where ${\bf u_{\it i}}$ is the local soft mode in unit cell
$i$; $\{ { \bf v_{\it i}} \}$ are the dimensionless local displacements
which are related to the inhomogeneous strain variables inside each
cell \cite{ZhongDavid};
$\eta_{\it H}$ is the homogeneous strain tensor; and the
$\{ \sigma_{{\it j}}\}$ characterize the atomic configuration of the
alloy.  That is,  $\sigma_{\it j}$=+1 or $-1$ corresponds to the
presence of a B$'$ or B$''$ atom, respectively, at lattice site $j$
of the A(B$'$$_{1-x}$B$''$$_{x}$)O$_{3}$ alloy.  The energy $E_{\rm ave}$
depends only on the soft mode and strain variables.  The $\{
\sigma_{{\it j}}\}$ parameters are thus incorporated into the
second energy term $E_{\rm loc}$.

For $E_{\rm ave}$, we generalize the analytical expression
successfully used in Ref.~\cite{ZhongDavid} for {\it simple}
ABO$_{3}$ systems to the case of an alloy.  This generalization
simply consists in using the virtual crystal approximation (VCA)
\cite{VCA}, i.e., in replacing A(B$'$$_{1-x}$,B$''$$_{x}$)O$_{3}$ by
a uniform but composition-dependent ``virtual'' ABO$_{3}$ system.
$E_{\rm ave}$ thus consists of five parts:  a local-mode
self-energy, a long-range dipole-dipole interaction, a short range
interaction between soft modes, an elastic energy, and an
interaction between the local modes and local strain
\cite{ZhongDavid}.

For  $E_{\rm loc}$, we propose an expression that includes (i) the
{\it onsite} effect of alloying on the self-energy up to the fourth
order in the local-mode vector $\bf{u_{\it i}}$; and (ii) {\it intersite}
contributions which are linear in  $\bf{u_{\it i}}$ and in
$\bf{v_{\it i}}$:
\begin{eqnarray}
  E_{\rm loc} && (\{ { \bf u_{\it i}} \},\{ { \bf v_{\it i}} \},
  \{ \sigma_{\it j} \}) = \nonumber \\
  && \sum_{i} [ \Delta \alpha (\sigma_{\it i}) ~ u_{\it i}^{4} ~+
  ~ \Delta \gamma (\sigma_{\it i}) ~( u_{\it ix}^{2}u_{\it iy}^{2} +
  u_{\it iy}^{2}u_{\it iz}^{2} + u_{\it iz}^{2} u_{\it ix}^{2})]\nonumber \\
  && +~ \sum_{ij}
  [Q_{\it j,i}(\sigma_{\it j}) ~ { \bf e_{\it ji}} \cdot { \bf u_{\it i}}~+~
  R_{\it j,i}(\sigma_{\it j}) ~ { \bf f_{\it ji}} \cdot { \bf v_{\it i} })]
  \;\;,
\end{eqnarray}
where the sums over $i$ and $j$ run over unit cells and
mixed sublattice sites, respectively.  $\Delta \alpha
(\sigma_{\it i})$ and $\Delta \gamma (\sigma_{\it i})$ characterize
the onsite contribution of alloying, while $Q_{{\it
j,i}}(\sigma_{\it j})$ and $R_{{\it j,i}}(\sigma_{\it j})$ are
related to alloying-induced intersite interactions.
Here $\bf{e_{\it ji}}$ is a unit vector
joining the site $j$ to the center of the soft-mode vector $\bf{u_{\it
i}}$, and $\bf{f_{\it ji}}$ is a unit vector joining the site $j$ to the
origin of the displacement $\bf{v_{\it i}}$.  In principle, terms involving
higher powers of $\bf{u_{\it i}}$ and $\bf{v_{\it i}}$ might be
included to improve the quality of the expansion, but as will
be shown below, we find this level of expansion to give a very good
account of experimental findings.  We also find that  $Q_{\it
j,i}(\sigma_{\it j})$ and $R_{\it j,i}(\sigma_{\it j})$ decrease
rapidly as the distance between $i$ and $j$ increases. As a
result, we included contributions up to the third neighbors for
$Q_{\it j,i}(\sigma_{\it j})$, and over the first-neighbor shell
for $R_{\it j,i}(\sigma_{\it j})$.  

All the parameters of Eqs.~(1) and (2) are derived from
first principles.  The 18 parameters of $E_{\rm ave}$ (see Table II
of Ref.~\cite{ZhongDavid}) are determined by fitting the results of
first-principles VCA calculations.  On the other hand, $\Delta
\alpha (\sigma_{\it i})$, $\Delta \gamma (\sigma_{\it i})$,
$Q_{{\it j,i}}(\sigma_{\it j})$ and $R_{{\it j,i}}(\sigma_{\it
j})$ are derived by performing first-principles calculations in
which a true atom [e.g., Ti or Zr in Pb(Zr,Ti)O$_{3}$] is
surrounded by VCA atoms.  The first-principles method used in the
present study is the plane-wave ultrasoft-pseudopotential method
\cite{USPP} within the local-density approximation (LDA)
\cite{LDA}. The VCA approach adopted averages the 
B$'$ and B$''$ pseudopotentials, and is the one of
Ref.~\cite{LaurentDavid3}.

Once our effective Hamiltonian is fully specified, the total energy
of Eq.~(1) is used in Monte-Carlo simulations to compute {\it
finite-temperature} properties of ferroelectric alloys.  We
typically use a $12\times 12\times 12$ supercell
(8640 atoms), since this choice yields well-converged
results.  The $\{ \sigma_{\it j} \}$ variables are chosen randomly
in order to mimic maximal compositional disorder -- consistent
with experimental reality \cite{Cross}-- and are kept fixed during the
Monte-Carlo simulations.  We find that averaging our results over a
couple of different realizations of the disorder leads to
well-converged statistical properties.  The outputs of the Monte-Carlo
procedure are the local mode vectors ${\bf u}$ (directly
related to the electrical polarization), and the homogeneous
strain tensor $ \eta_{\it H}$.  We use the correlation-function
approach of Ref.~\cite{Alberto1} to derive the piezoelectric response
from these Monte-carlo simulations.  Up to 10$^{6}$ Monte-Carlo
sweeps are first performed to equilibrate the system, and then
2$\times$10$^{4}$ sweeps are used to get the various
statistical averages.  The temperature is decreased in small
steps.

\begin{figure}
\epsfxsize=2.8 truein
\epsfbox{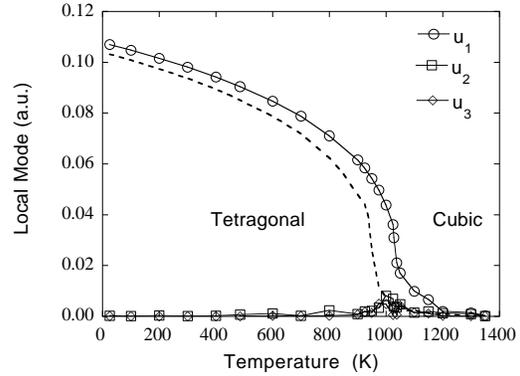}
\vskip 0.15truein
\caption{Largest, middle, and smallest average cartesian
coordinates $u_{1}$, $u_{2}$ and $u_{3}$ of the local-mode vector
as a function of temperature  in disordered single crystals of
Pb(Zr$_{0.5}$Ti$_{0.5}$)O$_{3}$.  Solid lines denote inclusion, while dashed li
nes denote
neglect, of $E_{\rm loc}$ in Eq.~(1).  For clarity, $u_{2}$ and
$u_{3}$ are not shown for the latter (VCA) case; they
are nearly zero at all temperatures.}
\end{figure}

Figure 1 shows the largest, middle and smallest cartesian
coordinates ($u_{1}$, $u_{2}$ and $u_{3}$) of the supercell
average of the local mode vectors in
Pb(Zr$_{0.5}$Ti$_{0.5}$)O$_{3}$ as a function of the temperature,
as predicted by our approach described by Eqs.~(1) and (2). Each
coordinate is close to zero at high temperature, characterizing a
paraelectric cubic phase.  As the system is cooled down, $u_{1}$
drastically  increases while $u_{2}$ and $u_{3}$ remain nearly null.
This indicates a transition to a ferroelectric
tetragonal phase, consistent with measurements \cite{Lines}.  We
predict that the spontaneous polarization reaches 0.79 C/m$^{2}$ at
very low temperature, which compares well with the
first-principles results of 0.70 and 0.74 C/m$^{2}$ \cite{Szabo1}.
The tetragonal axial ratio $c/a$ ranges from 1 close to the
transition region to 1.02 for lower temperature. This is  in good
agreement  both  with the experimental value of 1.02 -- 1.025
\cite{Lines,Noheda1} obtained for disordered samples, and with the
first-principles result of 1.03 obtained for an ordered alloy
\cite{LaurentDavid1}.  Figure 1 also shows the predictions of the
VCA alloy theory, corresponding to the neglect of $E_{\rm loc}$ in
Eq.~(1).  Interestingly, one sees that $E_{\rm loc}$ has no
effect on the phase transition {\it sequence}, which is consistent
with recent findings that the VCA approach can  reproduce some
structural properties of PZT \cite{LaurentDavid3,Rappe}.
Whether or not  $E_{\rm loc}$ is included in the total energy,
we find a Curie temperature $T_{c}$ that is higher than the
experimental value of 640 K \cite{YamamotoJJ}. This difficulty of
reproducing $T_{c}$ is a general feature of the effective-Hamiltonian
approach \cite{ZhongDavid,Rabe,Krakaeur}, and may be
due to higher perturbative terms neglected in the analytical
expression for the total energy.  In order to compare our
results with experimental data, we will henceforth rescale our
temperature as in Ref.~\cite{Alberto1} so that the theoretical
$T_c$ is forced to match the experimental one.

\begin{figure}
\epsfxsize=2.8 truein
\epsfbox{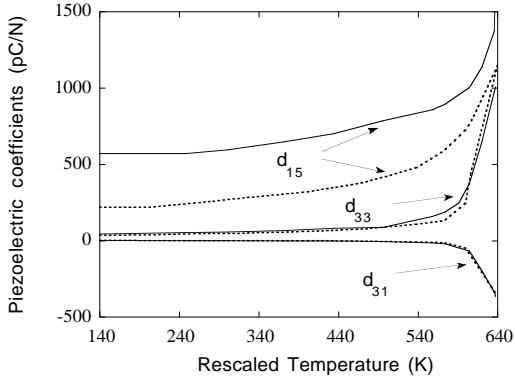}
\vskip 0.15truein
\caption{Piezoelectric coefficients as a function of temperature in
disordered (4mm) Pb(Zr$_{0.5}$Ti$_{0.5}$)O$_{3}$.
 Solid lines denote inclusion, while dashed
lines denote neglect, of $E_{\rm loc}$ in Eq.~(1).  Temperature has
been rescaled to fit the experimental value of the Curie
temperature. Statistical errors are estimated to be $\sim$10\% of
the values displayed.}
\end{figure}

Figure 2 shows the piezoelectric coefficients predicted for a
tetragonal single crystal of Pb(Zr$_{0.5}$Ti$_{0.5}$)O$_{3}$ as a
function of the rescaled temperature, when neglecting or
incorporating $E_{\rm loc}$ in Eq.~(1).  The independent
coefficients for the 4mm point group are $d_{33}$, $d_{31}$ and
$d_{15}$.  One can notice that inclusion of $E_{\rm loc}$ has only a
small effect on $d_{31}$ and $d_{33}$: $d_{31}$ is rather small for
any temperature except near the transition, and $d_{33}$ is
around 50 -- 55 pC/N at room temperature in both simulations.  Using
the experimental values of the elastic compliances \cite{Szabo1} to
compute $e_{33}$ from our calculated $d_{31}$ and $d_{33}$, we find
that our alloy effective hamiltonian leads to an $e_{33}$ of 4.3
C/m$^{2}$, while neglecting  $E_{\rm loc}$ in Eq.~(1) yields a
similar $e_{33}$ of 3.8 C/m$^{2}$ at low temperature.  Both
predictions agree well with the first-principles results ranging
between 3.4 and 4.8  C/m$^{2}$
\cite{Szabo1,LaurentDavid1,LaurentDavid3}, confirming that the
VCA can reproduce the $e_{33}$ coefficient of PZT
\cite{LaurentDavid3}.

Figure 2 also demonstrates that incorporating $E_{\rm loc}$ in the
total energy leads to a {\it large enhancement of the $d_{15}$
coefficient}, which is consistent with recent measurements revealing that
the piezoelectric elongation of the tetragonal unit cell of PZT does not occur
along the polar direction \cite{Noheda3}. This enhancement 
is highly relevant for the piezoelectric response
$d_{33}$ in ceramic samples, denoted $d_{33,c}$, which
involves an average of the form
\begin{equation}
d_{33,c}~=~\int_{0}^{\pi/2}
[(d_{31}+d_{15})\sin^2\theta+d_{33}\,\cos^2\theta]
\sin\theta\,\cos\theta\,d\theta
\end{equation}
over the single-crystal coefficients.  The true alloy approach
of Eqs.~(1)-(2) leads to a $d_{33,c}$ of 163 pC/N at
room temperature, in excellent agreement with the experimental
value of 170 pC/N \cite{Du1,Berlincourt}.  On the other hand,
neglecting $E_{\rm loc}$ leads to a smaller $d_{33,c}$ of 90 pC/N.
This difference clearly demonstrates the necessity of incorporating the
local alloying effect into the total energy for understanding the
large piezoelectric response of PZT ceramics near the MPB.

\begin{figure}
\epsfxsize=2.8 truein
\epsfbox{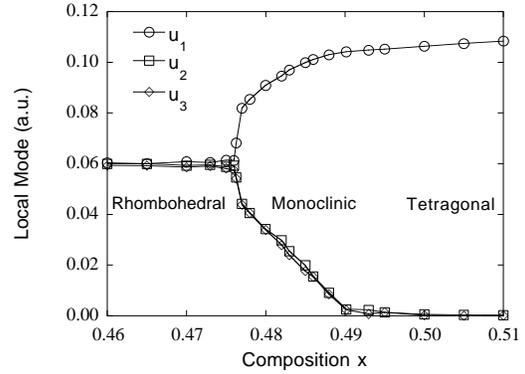}
\vskip 0.15truein
\caption{Similar to Fig.~(1), but computed as a function
of composition $x$.  The temperature of the simulation is 50 K,
corresponding to a rescaled experimental value of 30 K (see text).}
\end{figure}

We now use our alloy effective Hamiltonian to investigate the
low-temperature phases of  Pb(Zr$_{1-x}$Ti$_{x}$)O$_{3}$ near the
MPB.  We choose a constant temperature of 50 K in the Monte-Carlo
simulations, and vary the overall composition of the solid
solution.  This compositional variation affects two quantities:
(i) the populations of $\sigma_{\it j}$ equal to $+1$ or $-1$, and
(ii) the alloy-related
parameters.  For the latter, only the parameters entering the
local-mode self-energy of $E_{\rm ave}$ \cite{ZhongDavid}, and the
$\Delta \alpha (\sigma_{\it i})$ and $\Delta \gamma (\sigma_{\it
i})$ in Eq.~(2), are allowed to be composition-dependent.  This
composition-dependence is assumed to be linear, and is determined
by performing first-principles simulations on two different
compositional cells.  Such a linear composition-dependence approach
is only realistic when exploring a narrow range of compositions, as
done in the present study.

Figure 3 shows that the local mode, and hence the polarization, is
parallel to the pseudo-cubic [001] direction for Ti compositions
larger than 50\%, which is consistent with a tetragonal phase.
For compositions lower than 47\%, the polarization becomes
parallel to the pseudo-cubic [111] direction, indicating the
''high temperature'' rhombohedral phase \cite{Lines,Note}. 
 The most interesting feature of Fig.~3 is the
behavior of the local mode for  the compositional range between
47.5\% and 49.5\%: as $x$ decreases, $u_{1}$ decreases, while
$u_{2}$ and $u_{3}$ increase and remain nearly equal to each other.
This behavior is characteristic of an intermediate phase that is neither
tetragonal
nor rhombohedral.  The strain tensor given by our simulations
indicates that this intermediate phase is the monoclinic phase
experimentally found by Noheda {\it et al}. We further predict that
the monoclinic phase for $x \simeq 48\%$ can be characterized by
an angle of 90.7$^\circ$ and by lattice vectors
{\bf a$_{m}$} = $a_{0}$($-$1.005,$-$1.005,$-$0.009),
{\bf b$_{m}$} = $a_{0}$(1.002,$-$1.002,0.000), and
{\bf c$_{m}$} = $a_{0}$(0.004,0.004,1.018),
where a$_{0}$ is a cubic lattice constant.  All these predictions
are in excellent quantitative agreement with the experimental
results of Ref.~\cite{Noheda1}.  Figure 3 clearly demonstrates that
the monoclinic phase acts as a bridge  between the rhombohedral and
tetragonal phases, as indicated by the continuous rotation of
the polarization as a function of composition. 
Our computational scheme
is also able to reproduce the compositional range narrowing of
the monoclinic phase observed when increasing the temperature \cite{Noheda3}.
It should be
noted that a VCA-only calculation (i.e., neglecting $E_{\rm loc}$)
does not reveal a monoclinic phase.  This finding demonstrates once
again the need for incorporating the local effect of alloying into
the total energy to study subtle effects.

In summary, we have developed a first-principles derived
computational scheme to study finite-temperature properties of
Pb(Zr$_{1-x}$Ti$_{x}$)O$_{3}$ solid solutions near the MPB as a
function of composition and temperature. We find that there is a
low-temperature monoclinic phase acting as a bridge between the
rhombohedral phase, existing for $x < 0.47$, and the tetragonal phase,
occurring for $x$ larger than 0.50.  The predicted structural data
are in very good agreement with measurements, as well as with
direct first-principles calculations.  The use of this approach
also provides an explanation for the large experimental value of
$d_{33}$ in tetragonal ceramics of PZT near the MPB
\cite{Berlincourt}.  This large piezoelectricity is simply due to
the very large value of the $d_{15}$ coefficient predicted to occur
in the single crystal.

L.B.~thanks the financial assistance provided by the Arkansas
Science and Technology Authority (grant N99-B-21), and
the National Science Foundation (grant DMR-9983678).
A.G.\ acknowledges support from the Spanish Ministry of Education
(grant PB97-0598).
D.V.\ acknowledges
the financial support of Office of Naval Research grant N00014-97-1-0048.
We wish to thank B. Noheda,  H. Chen, M. Cohen, E. Cross, T. Egami
and Q. Zhang for very useful discussions.

\newpage

\narrowtext


\begin{references}
\bibitem{Note}  Our model cannot predict the
``low-temperature'' rhombohedral phase \cite{Lines} since this involves
 octahedra-tilting degrees of freedom that are not including in the present 
model.  Our simulations thus also demonstrate that the oxygen tilts are not 
the driving force for the transition or for the occurrence of a monoclinic 
phase.
\end{references}
\end{document}